\documentclass[twocolumn]{aastex631}

\usepackage{amsmath}
\usepackage{physics}
\usepackage[caption=false]{subfig}

\begin{document}

\title{Semi-supervised Learning for Detecting Inverse Compton Emission in Galaxy Clusters}

\author{Sheng-Chieh Lin}
\affiliation{Department of Physics and Astronomy, University of Kentucky, Lexington, KY, USA}

\author{Yuanyuan Su}
\affiliation{Department of Physics and Astronomy, University of Kentucky, Lexington, KY, USA}

\author{Fabio Gastaldello}
\affiliation{INAF, Istituto di Astrofisica Spaziale e Fisica Cosmica di Milano, Via A. Corti 12, 20133 Milano, Italy}

\author{Nathan Jacobs}
\affiliation{Department of Computer Science \& Engineering, Washington University in St. Louis, St. Louis, MO, USA}

\begin{abstract}
Inverse Compton (IC) emission associated with the non-thermal component of the intracluster medium (ICM) has been a long sought phenomenon in cluster physics. 
Traditional spectral fitting often suffers from the degeneracy between the two-temperature thermal spectrum (2T) and the one-temperature plus IC power-law spectrum (1T+IC).
We present a semi-supervised deep learning approach to search for IC emission in galaxy clusters. 
We employ a conditional autoencoder (CAE), which is based on an autoencoder with latent representations trained to constrain the thermal parameters of the ICM.
The algorithm is trained and tested using synthetic NuSTAR X-ray spectra with instrumental and astrophysical backgrounds included.
The training data set only contains 2T spectra, which is more common than 1T+IC spectra.
Anomaly detection is performed on the validation and test datasets, consisting of 2T spectra as the normal set and 1T+IC spectra as anomalies.
With a threshold anomaly score, chosen based on cross-validation, our algorithm is able to identify spectra that contain an IC component in the test dataset, with a balanced accuracy (BAcc) of $0.64$, which outperforms traditional spectral fitting (BAcc $=0.55$) and ordinary autoencoder (BAcc $=0.55$). 
Traditional spectral fitting is better at identifying IC cases among true IC spectra (a better recall), while IC predictions made by CAE have a higher chance of being true IC cases (a better precision),
demonstrating their mutual complement to each other.
\end{abstract}


\section{Introduction}

Clusters of galaxies are the largest gravitational bound and multi-component objects with masses of above $10^{14} M_{\odot}$, and gas temperatures of a few keV \citep[$\sim 10^8$\,K, see][for a review]{2022hxga.book...93Z}. They are the end result of the hierarchical structure formation in the framework of $\Lambda$CDM. 
Throughout cosmic time, clusters accumulate matter from the cosmic web and undergo multiple merging events (e.g., \citealt{1991ApJ...379...52W}; \citealt{2005Natur.435..629S}).
Understanding the astrophysics operating within clusters is central to utilizing galaxy clusters for cosmology \citep{2020MNRAS.498.5620S,2022MNRAS.512.3885L}. 
In clusters, a significant portion of baryonic matter ($\gtrsim 90\%$) resides in the intracluster medium (ICM) which constitutes the hot ionized gas phase that radiates in X-rays mainly through thermal Bremsstrahlung.

The ICM is a weakly magnetized plasma in which magnetic fields play an important role in suppressing transport processes (\citealt{2010ApJ...713.1332R,2016ApJ...821...40S,2017ApJ...834...74S,2019AJ....158....6S,2023MNRAS.526.6052S}).
Cluster magnetic fields that interact with relativistic electrons produce diffuse synchrotron emission, resulting in prominent phenomena known as radio halos in the centers of clusters and radio relics in the periphery \citep[see][for a review]{2019SSRv..215...16V}.
The estimate of magnetic field strength through the synchrotron emission would rely on assumptions including the equipartition of the cosmic ray particles and the cluster magnetic field.
One can also derive the field strength through the Faraday rotation (FR) of radio galaxies within or behind the clusters. However, it would depend on the density profile of the gas and the local geometry of the magnetic fields.
Inverse Compton (IC) emission at hard X-rays can offer a direct and unbiased measure of cluster magnetic fields.
In theory, the same relativistic electrons that produce the radio emission can up-scatter the cosmic background photons via the Inverse Compton mechanism, producing the non-thermal emission in hard X-ray \citep{2015A&A...582A..20B}.
The volume-average magnetic field strength is directly linked to the ratio of synchrotron flux to IC flux.

Probing the IC emission requires instruments that are sensitive to the hard X-ray regime, covering the energy range to $\sim 20$ keV and above, to minimize the emission from the thermal content of galaxy clusters.
Early attempts to detect IC emission date back to the HEAO-1 observatory, which placed upper limits on the IC flux, corresponding to a lower limit of $0.1$ $\mu$G of the magnetic field, averaging over a few clusters \citep{1987ApJ...320..139R, 1988ApJ...333..133R}.
The exploration was furthered carried on with RXTE and Beppo-SAX \citep[for a review, see][]{2008SSRv..134...71R}, with many claimed detections being either marginal or controversial \citep{2004ApJ...602L..73F, 2004A&A...414L..41R, 2007ApJ...654L...9F}.
The inconclusive results led to subsequent investigations with \textit{swift} and \textit{Suzaku}, which did not confirm the previously claimed IC detections
\citep{2012ApJ...748...67W, 2012RAA....12..973O, 2014A&A...562A..60O}.

The Nuclear Spectroscopic Telescope Array \citep[NuSTAR;][]{2013ApJ...770..103H} mission is the first of its kind to spatially resolve cluster emissions at hard X-ray bands above $10$ keV (range: $3-80$ keV).
=The NuSTAR observatory carries two co-aligned identical telescopes, each with an angular resolution of $58''$ (HPD; $18''$ FWHM) and with an energy resolution of $0.4$ keV at $6$ keV.
The focal plane size of NuSTAR is relatively small with a field-of-view (FOV) of $12'\times12'$.
As a first attempt to search for the IC emission using NuSTAR, \citealt{2014ApJ...792...48W} reports the results on the Bullet cluster in which the detection of IC cannot be made.
More specifically, the two-temperature plasma model can fully explain the observed spectra of the Bullet cluster, hence the non-thermal component is not required.
Later on, many efforts have been made based on either the standalone NuSTAR data, or joint analysis of the NuSTAR and other observations, as listed below.
\citealt{2015ApJ...800..139G} put an upper limit of $5.1\times10^{-12}$ erg cm$^{-2}$ s$^{-1}$ on the IC flux of the Coma cluster using NuSTAR.
\citealt{2019A&A...628A..83C} jointly analyze the data from the NuSTAR and XMM-Netwon observations of the Abell 523 cluster, reaching an upper limit of $3.17\times10^{-14}$ erg cm$^{-2}$ s$^{-1}$ in a local region around the radio halo.
\citealt{2021ApJ...906...87R} present a search of the IC emission on the Abell 2163 cluster, using NuSTAR, resulting in an upper limit of $4.0\times10^{-12}$ erg cm$^{-2}$ s$^{-1}$.
Following the previous work, \citealt{2023ApJ...954...76R} report the search for IC emission in two more clusters, Abell 665 and Abell 2146, which have upper limits of $6.0\times10^{-13}$ erg cm$^{-2}$ s$^{-1}$ and $8.5\times10^{-13}$ erg cm$^{-2}$ s$^{-1}$, respectively.
\citealt{2023ApJ...942...79T} make use of both the NuSTAR and Chandra observations of the cluster CL\,0217+70 to search for IC signal and obtain an upper limit of $2.7\times10^{-12}$ erg cm$^{-2}$ s$^{-1}$.
\citealt{2024ApJ...962...94T} conduct a similar study on the cluster ZWCL\,1856.8 while no conclusive results are drawn due to insufficient photon counts.
\citealt{2023MNRAS.524.4939M} report a detection of IC signal based on the study of a low-mass galaxy group MRC 0116 +111 using the XMM-Newton Observatory.
Thanks to its low gas temperature of kT $=0.7-0.8$ keV, its power-law IC flux of $1.3 \pm 0.28 \times10^{-15}$ erg cm$^{-2}$ s$^{-1}$ is detected at $4.6 \sigma$, corresponding to a magnetic field of $1.9\pm0.3$ $\mu$G.
However, the non-thermal emission detected in their work is likely associated with the radio lobes of central galaxy.

Merging clusters represent the most energetic phenomenon in the Universe, manifested in both X-ray and radio observations.
However, as shown above, the search for the IC signal in galaxy clusters has been challenging, as observations often yielded only upper limits.
This is primarily due to the relatively faint IC emission and the difficulty in accurately modeling the complicated hot ICM and background spectra. 
Nevertheless, NuSTAR is the primary current (and foreseeable future) observatory with the capability of focusing hard X-rays, making NuSTAR 
observations our best chance to detect IC emission in clusters.
In this work, we aim to build a machine learning (ML) model to identify the IC signal from the hard X-ray spectra from NuSTAR, complementing traditional spectral analysis. 
We could have treated it as a supervised task and provided labels for both IC and non-IC simulated spectra. Supervised learning is known to outperform unsupervised or semi-supervised learning given the same amount of labeled training data. However, particularly for observational astronomy, the availability of labeled data in the real world is severely lacking. In our case, known real IC spectra are (and will be) much more rare compared to known real thermal spectra. Our vision and motivation are to design a prototype model that has the potential to be trained on real data. 
In this perspective, an unsupervised or semi-supervised algorithm would be a more far-reaching choice and would inspire other areas of astronomy to best utilize the availability of large amount of unlabeled data and small number of labeled data.

We treat it as an anomaly detection task as IC is a rare phenomenon in nearby clusters. 
An autoencoder (AE) is adopted as the backbone algorithm.
It is a specific class of unsupervised learning algorithms that consists of two neural networks, an encoder and a decoder.
The encoder is tasked with extracting the latent representations from the input data, while the decoder uses the representations to regenerate the inputs.
Typically, latent representations have a lower dimension than the input data.
Pre-trained autoencoders are often reused for other tasks, such as anomaly detection, or various downstream applications (e.g., \citealt{2019MNRAS.487L..24T}; \citealt{2020MNRAS.494.3750C}; \citealt{2022NatPh..18..112G}).
For our purpose, we implement a conditional autoencoder (CAE) -- a semi-supervised algorithm based on the original AE. The algorithm is trained and tested on synthetic NuSTAR data.
The structure of this paper is as follows.
We describe data preparation in Sec.~\ref{sec:data}.
We introduce the algorithm and briefly discuss anomaly detection in Sec.~\ref{sec:model}. In Sec.~\ref{sec:results}, we compare the performance of our algorithm and traditional spectral fitting, and discuss the observed patterns of correctly identified cases. Finally, our work is summarized in Sec.~\ref{sec:summary}.

\section{NuSTAR Synthetic Spectra}
\label{sec:data}

Training deep learning models requires a large amount of data to acquire predictive capabilities.
Given the limited number of clusters observed by NuSTAR, we used the fast simulation tool \textsc{fakeit} from the X-ray data analysis package XSPEC \citep{1996ASPC..101...17A} to simulate a sufficient number of synthetic spectra.
In the following, we describe the data-generating processes, including the source model and background components.

The dataset we prepared for training is based on a 2T model -- a thermal plasma model with two-temperature components absorbed by the foreground Galactic gas -- \texttt{tbabs*(apec+apec)}.
For the validation and test datasets, in addition to the 2T spectra, we include an alternative 1T+IC model -- \texttt{tbabs*(apec+po)}, which describes a thermal plasma component and an IC component with a power-law energy distribution.

To produce realistic spectra, response files from the NuSTAR observation of the Coma cluster \citep{2015ApJ...800..139G} are used and convolved with the thermal/non-thermal models.
For the \texttt{apec} component in all datasets, values of the thermal parameters are randomly generated from $kT= 1.00$--$10.00$ keV (temperature); $Z=0.10$--$1.00$ $Z_{\odot}$ (metallicity); $z=0.01$--$0.10$ (redshift); $n=0.01$--$1.00$ (normalization).
For 2T spectra, the two thermal components \texttt{apec+apec} have their redshift and metallicity linked, while the temperature and normalization are left to vary freely.
For the IC component, the photon index of the power law is fixed at $2$ and the normalization is set to vary from $2\times10^{-4}$ to $2\times10^{-3}$, where the upper bound corresponds to the upper limit of the IC flux of the Coma cluster \citep{2015ApJ...800..139G}.
The hydrogen column density of the galactic absorption term \texttt{tbabs} is fixed at $N_{H}=8.58\times10^{19}$ cm$^{-2}$ for all simulated spectra as for the Coma cluster. 

\begin{figure}
\centering\includegraphics[width=8cm]{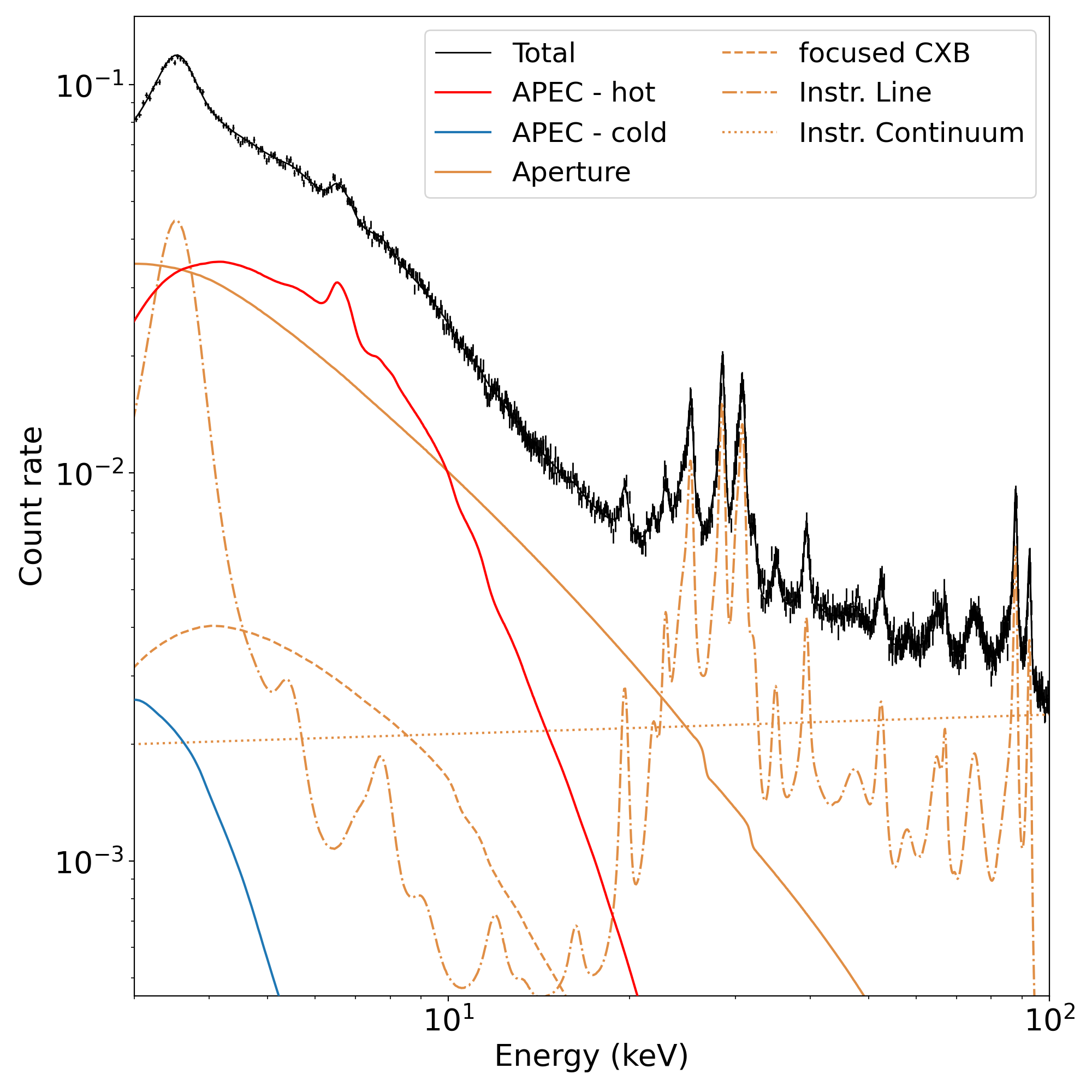}

\caption{
        An example of synthetic NuSTAR spectrum consists of thermal emission and four background models.
        The synthetic spectrum shown as black points is generated using \textsc{fakeit} for a (source plus background) model shown as black curve.
        The cluster emission model is adopted from the thermal parameters of the Coma cluster, determined from the best-fit of an absorbed two-temperature model of \texttt{tbabs*(apec+apec)} \citep{2015ApJ...800..139G}.
        The red spectrum (APEC) is the hotter component at $9.03$ keV, and the blue spectrum is the cooler one at $1.02$ keV.
        Several background components (brown) are taken into account including instrumental continuum background (Instr. Continuum, dotted), focused cosmic X-ray background (focused CXB, dashed), aperture background (Aperture, solid), and instrumental lines and solar reflected component (Instr. Line, dashdotted).
}

\label{fig:spec}
\end{figure}

Realistic NuSTAR background components are considered in the construction of synthetic spectra.
We incorporate four main astrophysical backgrounds in our data generation procedure by superimposing the background models onto the source models.
Specifically, they are the instrumental continuum background from the Compton scatter of the high energy gamma rays, the aperture background from the cosmic X-ray background leaking through aperture stops, the instrumental emission lines together with the reflected solar X-rays in the soft bands, and the focused cosmic X-ray background.
These components are relatively stable and can be determined by empirically fitting spectra observed from source-free regions.
In our study, we use empirical models built from the blank-sky observations, which is also adopted in the analysis of the Coma cluster \citep{2015ApJ...800..139G}.
In Fig.~\ref{fig:spec}, we show an example of a 2T spectrum, in which various backgrounds clearly dominate the hard X-ray bands (E $>20$ keV), making the detection of faint IC emission more difficult.
Each synthetic observation is integrated for an exposure time of $1$\,Ms.
There are two identical detectors onboard NuSTAR.
We only considered one for simplicity, which would correspond to effectively less exposure time for NuSTAR.
The long exposure time we selected may be difficult to achieve, but we chose it to ensure it would not be hampered by low-count statisitics.
We generate $10000$ synthetic 2T spectra for training and validation, $1000$ 2T spectra plus $1000$ 1T+IC spectra for testing.
Only 2T spectra are included in the training (and cross-validation) dataset.
There is no overlap between the training dataset and test dataset.
Note that 1T spectra are included in the 2T training data set as there are cases in which 1T component strongly dominates the emission.
By construction, any non-thermal spectra, including those that are different from a 1T+IC model, will be identified as IC emission.
However, this algorithm is applied specifically to ICM where diffuse radio emission is detected. 
Our priori knowledge of the presence of relativistic electrons would strengthen the confidence of detecting an IC component for non-thermal spectra.

\section{Model}
\label{sec:model}

We implemented a conditional autoencoder to achieve a reconstruction task while utilizing the latent representatives to constrain the thermal parameters.
The code together with pretrained weights and test spectra are available on the github\footnote{the github repository \url{https://github.com/shengclin/iccae.git} under a CC-BY-4.0 License, and it is archived in Zenodo \citep{shengchieh_lin_2024_13738022}}.

\subsection{Autoencoders}


    \begin{figure*}
        \centering
        \subfloat[]{%
            \includegraphics[width=0.5\linewidth]{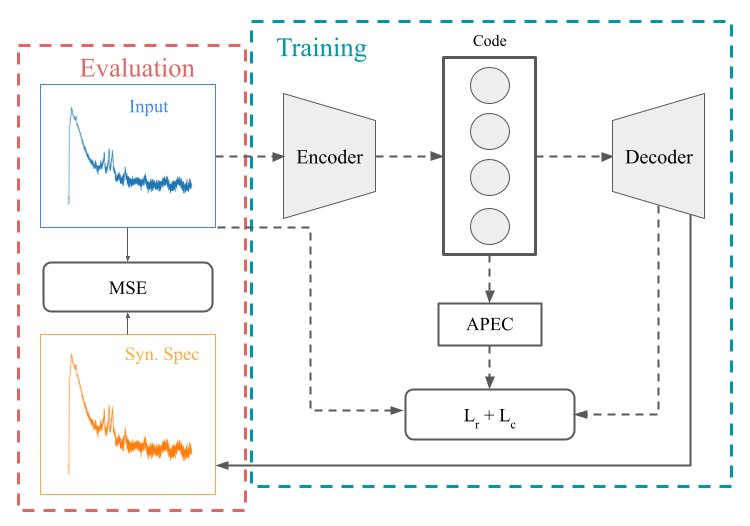}%
        }\subfloat[]{%
            \includegraphics[width=0.5\linewidth]{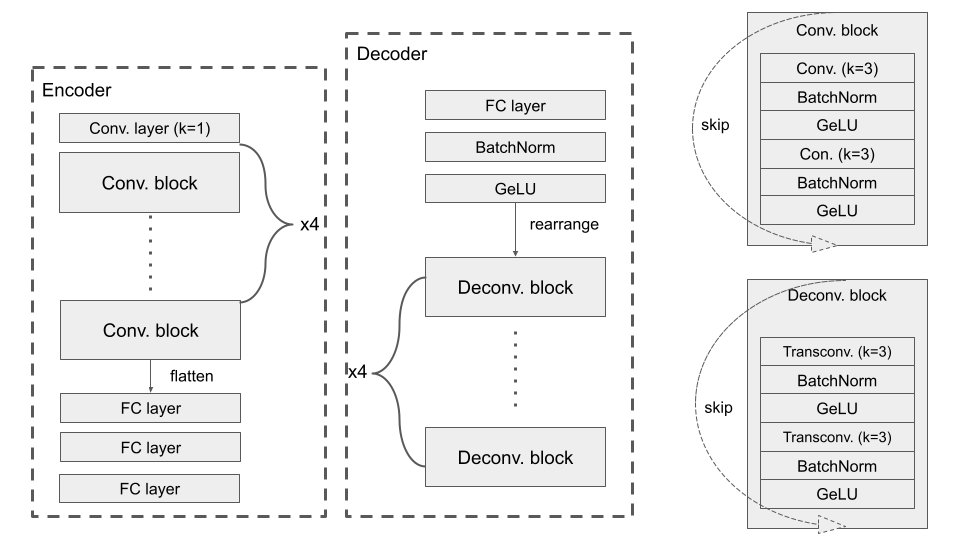}%
        }\caption{
            Schematic diagram of the algorithm. \textbf{(a)} Demonstration of the training and testing processes.
            The dashed arrows show the training work flow that ends in the round block of $L_r + L_c$, which represents the sum of the reconstruction loss and the constraining loss.
            The constraining loss, as shown in Eq. \ref{eq:loss}, is obtained by calculating the mean absolute error between the predicted values and the input values of the \texttt{apec} parameters (shown as the middle block labeled APEC).
            The solid arrows indicate the testing procedure in which we compare the input spectra with the reconstructed spectra using the mean squared error (MSE).
            \textbf{(b)} Detailed layouts of the encoder and decoder. Note that the skip mechanism is applied to both the convolutional and deconvolutional blocks.
        }
        \label{fig:architecture}
    \end{figure*}

A typical autoencoder consists of two neural networks (NNs), an encoder $\mathcal{F}$ and a decoder $\mathcal{G}$, with the ultimate learning goal of reconstructing the input data from the self-extracted representations \citep[or called latent representations;][]{4270182}.
The latent representations refer to the abstract or conceptual features of the data that are often not observable.
In the framework of autoencoder, the latent representations are the feature vectors output from the encoder, which can be used by the decoder to reconstruct the input data.
More specifically, the encoder $\mathcal{F}: \mathbb{R}^{l} \rightarrow \mathbb{R}^{d}$ learns to map the input data $x \in \mathbb{R}^{l}$ to the latent representations (the code) $u \in \mathbb{R}^{d}$, and the decoder $\mathcal{G}: \mathbb{R}^{d} \rightarrow \mathbb{R}^{l}$ outputs the reconstructed results $\tilde{x} \in \mathbb{R}^{l}$ using the representations $u$.
The autoencoder is trained by minimizing the loss function $\mathcal{L}$: arg min$_{F,G} E[\mathcal{L}(x, \mathcal{G}\circ\mathcal{F}(x))]$.
The dual networks in autoencoders are designed to extract useful information directly from the data itself without using any training labels.
In most cases with $d < l$, the bottleneck structure of autoencoders can effectively filter out the irrelevant features within data, distilling the latent representations for further reconstructions.
The bottleneck also helps prevent overfitting.

In addition to extracting useful information from nonlinear data, autoencoders are widely deployed in the tasks of anomaly detection \citep[see, e.g.][for a review]{PangGuansong2021DLfA}.
Anomalies, or outliers, are data from a distribution significantly different from \textit{normal} data \citep[see,][for a review]{AggarwalCharuC2016AItO}.
For the type of data that the autoencoders have not ``seen'' during the training phase, the compressed representations are expected to deviate from that of the \textit{normal} data (used in training), resulting in a higher reconstruction loss.
Following this concept, the simplest method to detect outliers is to monitor the reconstruction loss of the test sets.
Additionally, analysis of latent features can provide useful information regarding the distributions of both inliers and outliers if the latent representations remain complex or nonlinear.

\subsection{Conditional Autoencoder as an IC detector}

We build a conditional autoencoder, which learns from the spectra to attribute the latent representations to the properties of the thermal component.
CAE consists of a convolutional neural network as the encoder and a deconvolutional neural network as the decoder.
The model takes synthetic 2T spectra as input for training.
Thermal parameters of the 2T spectra is provided to the networks to constrain the latent variables via the loss function.
The output of CAE is the reconstruction of the input spectra.
Note that we utilize the transposed convolutional layer for deconvolutional operations.
Skip connection is implemented only in the encoder, resembling the ResNet architecture \citep{he2016deep}.
Each convolutional/deconvolutional layer is followed by a GeLU activation layer, and then batch-normalized through a mini batch with a size of $150$.
In total, there are $5$ stacks of convolutional and deconvolutional layers each in the encoder and decoder, respectively.
The layout of the whole architecture is shown in Fig. \ref{fig:architecture}.
The training of the model is implemented under the framework of PyTorch \citep{NEURIPS2019_9015}.

The total loss function consists of two components: mean squared error (MSE) loss, and mean absolute error (MAE) loss.
The MSE loss is used as the reconstruction loss ($L_r$) function, and the MAE loss as the constraining loss ($L_c$) for the latent representations.
Mathematically, the training loss function is designed as follows:
\begin{equation}
\label{eq:loss}
L = L_r+L_c=\frac{1}{N} \sum_{i=0}^{N} (y_i - \hat{y}_i)^2 + \frac{\epsilon}{N} \sum_{i=0}^{N} \abs{u_i^{(4)} - \theta_i}
\end{equation}
where $\epsilon$ is a constant fixed at $0.1$, $N$ is the number of spectra per batch, $y_i$ is the output vector of spectra from the decoder, $\hat{y}_i$ is the training vector, and $i$ is the index of a spectrum.
The $u_i^{(4)}$ is the latent representation, which is trained to predict the thermal parameters of the 2T spectra, $\theta_i = \{kT, Z, z, n\}_i$, for the ICM temperature, metallicity, redshift, and normalization, respectively.
The values of the thermal parameters and the input spectra themselves are provided as the ground truth in the training.
The ground truth temperature and normalization of the 2T spectra is obtained by fitting the training data, 2T spectra, with a one-temperature thermal model \texttt{tbabs*apec}.
Training labels for metallicity and redshift were set at their input parameter values when the spectra were generated.
Although the 1T model is a biased estimate of 2T spectra, the former is chosen as the ground truth because it allows the CAE to achieve better performance than using 2T parameters.
    
Training is performed on a single GPU card for $300$ epochs, which is sufficient for the training to converge.
Adam optimizer (\citealt{kingma2014adam}) with an initial learning rate of $lr=0.01$ is adopted and a plateau scheduler (\texttt{ReduceLROnPlateau}) is implemented to decrease the learning rate when the test loss hits the optimization plateau.
The CAE is trained and cross-validated on the 10,000 2T spectra in the training dataset.

\section{Results and discussions}
\label{sec:results}

Typical ICM spectra contain one or two-temperature thermal components. 
When it comes to IC detection, single-temperature ICM plus IC emission spectra cannot be easily distinguished from two-temperature components spectra, especially for massive clusters with high thermal temperature radiating in hard X-ray.
This degeneracy in the interpretation of the spectra presents the biggest challenge in identifying IC emission in galaxy clusters using NuSTAR observations.
Here, we present the result of using a CAE algorithm to distinguish 1T+IC spectra and 2T spectra.
Considering that IC emission is a relatively rare phenomenon in nature, we treat it as an anomaly detection process and do not use abnormal data for training.

\subsection{Performance evaluation of CAE}

\begin{figure}
        \centering
        \includegraphics[width=8cm]{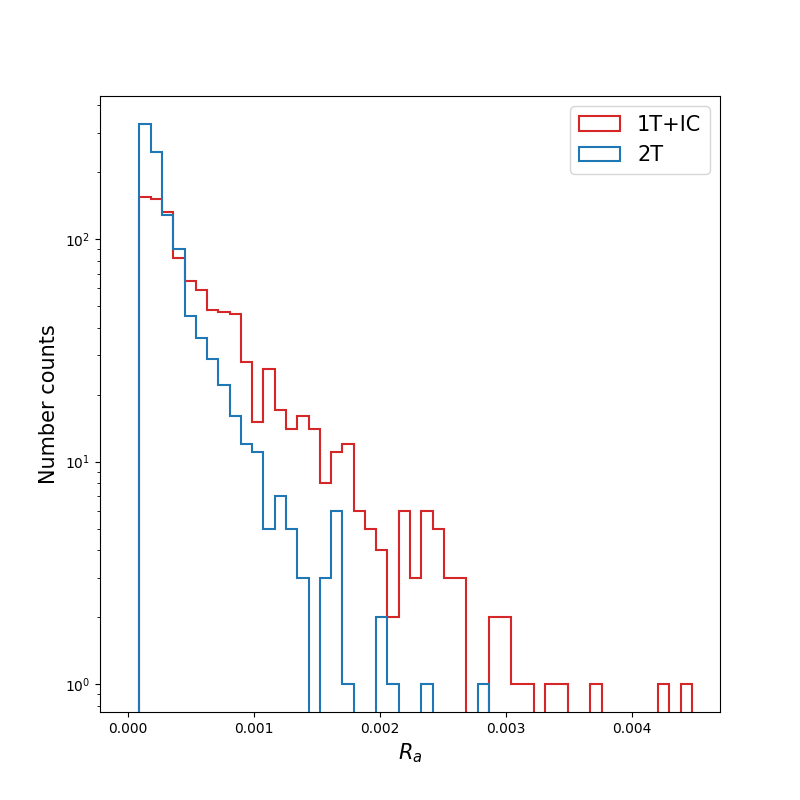}
        \caption{
            Distributions of the MSE (anomaly score $R_a$) of the spectra reconstruction for true 1T+IC (red) and true 2T (blue) spectra, returned by CAE.
        }
        \label{fig:anscore}
    \end{figure}

\begin{figure}
        \centering
        \includegraphics[width=8cm]{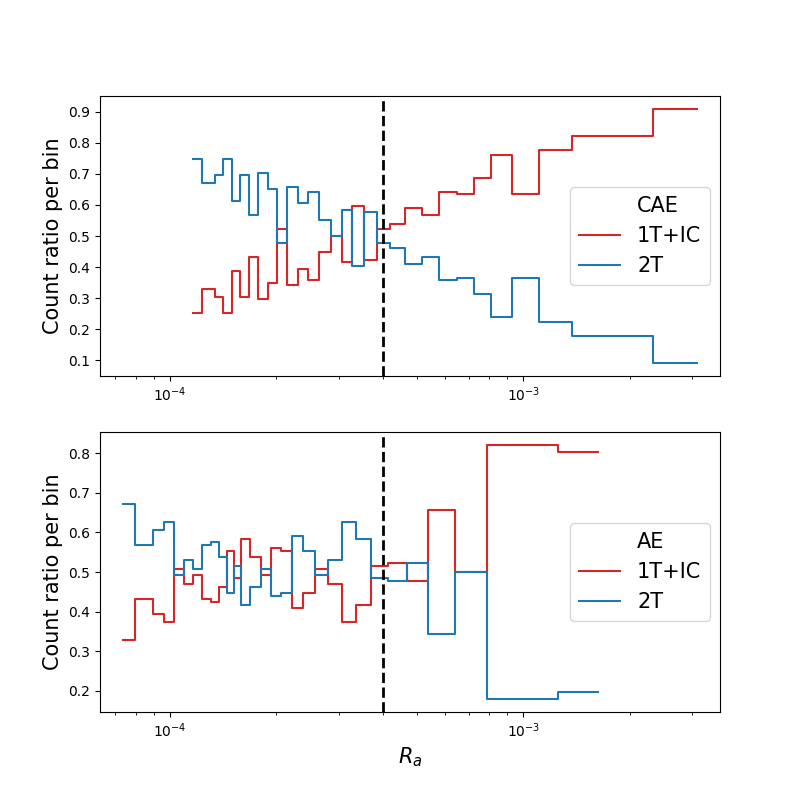}
        \caption{
            The spectrum number count as a function of anomaly score $R_a$.
            The vertical dashed line shows the threshold of $R_a=0.00042$ for classification decision making, above which spectra are classified as containing IC signal, while below which are classified as pure thermal spectra. True 2T and true 1T+IC data are marked in blue and red, respectively.
            Two panels show the results from two different autoencoders: \textit{upper} - CAE; \textit{lower} - ordinary autoencoder (AE).
        }
        \label{fig:acc}
    \end{figure}

To quantify the confidence in anomaly detection, we define an anomaly score, $R_{a}$, using the mean squared error (MSE) calculated from each pair of reconstructed spectra $S_r$ and the input ones $S_i$:
\begin{equation}
    R_{a} = \frac{1}{n_E} \sum_{E} (S_{r}(E)-S_{i}(E))^{2},
\end{equation}
where $n_E$ is the number of energy bins, for which we use a bin size of $\Delta E=0.04$\,keV over the energy range of $3-80$ keV.
When applied to the 1000 1T+IC plus 1000 2T spectra in the test dataset,
the resulting $R_{a}$ distribution is shown in Fig.~\ref{fig:anscore}.
The 1T+IC data have a high $R_{a}$ tail indicating bad reconstructions of some spectra.
To quantify the differences in $R_{a}$ between 2T and 1T+IC data, we examine the ratios of the number counts, which are shown in Fig.~\ref{fig:acc}.
For comparison, the results obtained from an ordinary autoencoder (AE) with the same layouts are also shown. CAE outperforms ordinary AE with a larger fraction of abnormal data among cases with high $R_a$.

\begin{figure}
        \centering
        \includegraphics[width=8cm]{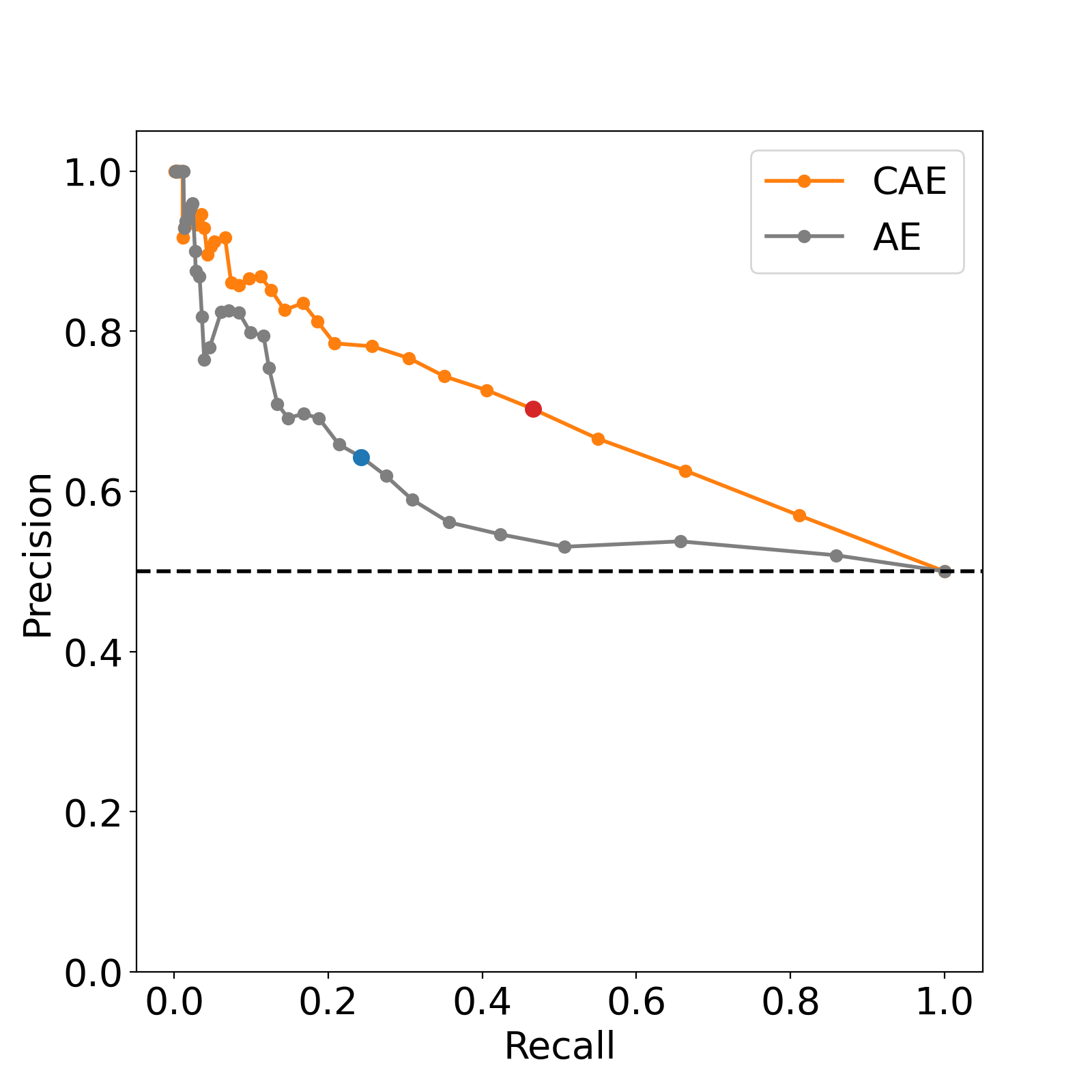}
        \caption{
            Relation between recall and precision.
            CAE is shown as orange, and the ordinary autoencoder (AE) is shown as grey.
            The red and blue points mark the values when using the chosen selection threshold of $R_a=0.00042$.
            The horizontal dashed line marks a precision of $0.5$, representing the performance of a random classifier. 
        }
        \label{fig:precision}
    \end{figure}

A threshold value of $R_a$ is needed for the purpose of classification.
Cases with a $R_a$ above and below the threshold value will be classified as 1T+IC and 2T spectra, respectively. 
One of the challenges regarding choosing a proper threshold comes from the trade-off between two statistical quantities, precision and recall.
For the sake of clarity, we will use $tp$, $fp$, $tn$, and $fn$ to represent the numbers of true positive, false positive, true negative, and false negative predictions, respectively.
Precision is defined as the ratio of true positives (correctly detected cases) to all positive predictions.
\begin{equation}
    {\rm Precision}=\frac{tp}{tp+fp}.
\end{equation}
Recall is defined as the fraction of true positives out of all positive cases. 
\begin{equation}
    {\rm Recall}=\frac{tp}{tp+fn}.
\end{equation}
The corresponding precision increases as recall decreases, when choosing different threshold values of $R_a$, as shown in Fig.~\ref{fig:precision} (PR curve).
Both CAE and AE are superior to a random classifier, regardless of the threshold. 

We choose balanced accuracy (BAcc) as the matrix for performance evaluation, which is the average of true positive
predictions divided by the number of positive cases and true
negative predictions divided by the number of negative cases:
\begin{equation}
{\rm BAcc}=\frac{1}{2}\left(\frac{tp}{tp+fn}+\frac{tn}{tn+fp}\right).
\end{equation}
Since IC is a rare phenomenon, the real-world dataset is likely to be dominated by thermal spectra. 
BAcc is chosen mainly because it is not biased by imbalanced datasets and it reflects a compromise between precision and recall.

We perform a $10$-fold cross-validation on the training dataset, containing 10000 2T spectra. 
CAE model is trained on 8000 2T spectra.
We select the best epoch that can minimize the validation loss for an additional 1000 spectra.
Then a threshold value of $R_a$ is chosen to maximize the BAcc for the remaining 1000 spectra.
We repeat this process 10 times to cycle through all the 10000 2T spectra in the training dataset and obtain 10 different threshold values, allowing us to choose the median of them as the final threshold value of $R_a=0.00042$.
With this threshold, CAE achieves a good balance between precision ($0.69$) and recall ($0.50$) in the test dataset, as shown in the PR curve in Fig.~\ref{fig:precision}, corresponding to a BAcc of $0.64$, while that of an ordinary AE is ${\rm BAcc}=0.55$.

\begin{figure}
    \centering
    \includegraphics[width=\columnwidth]{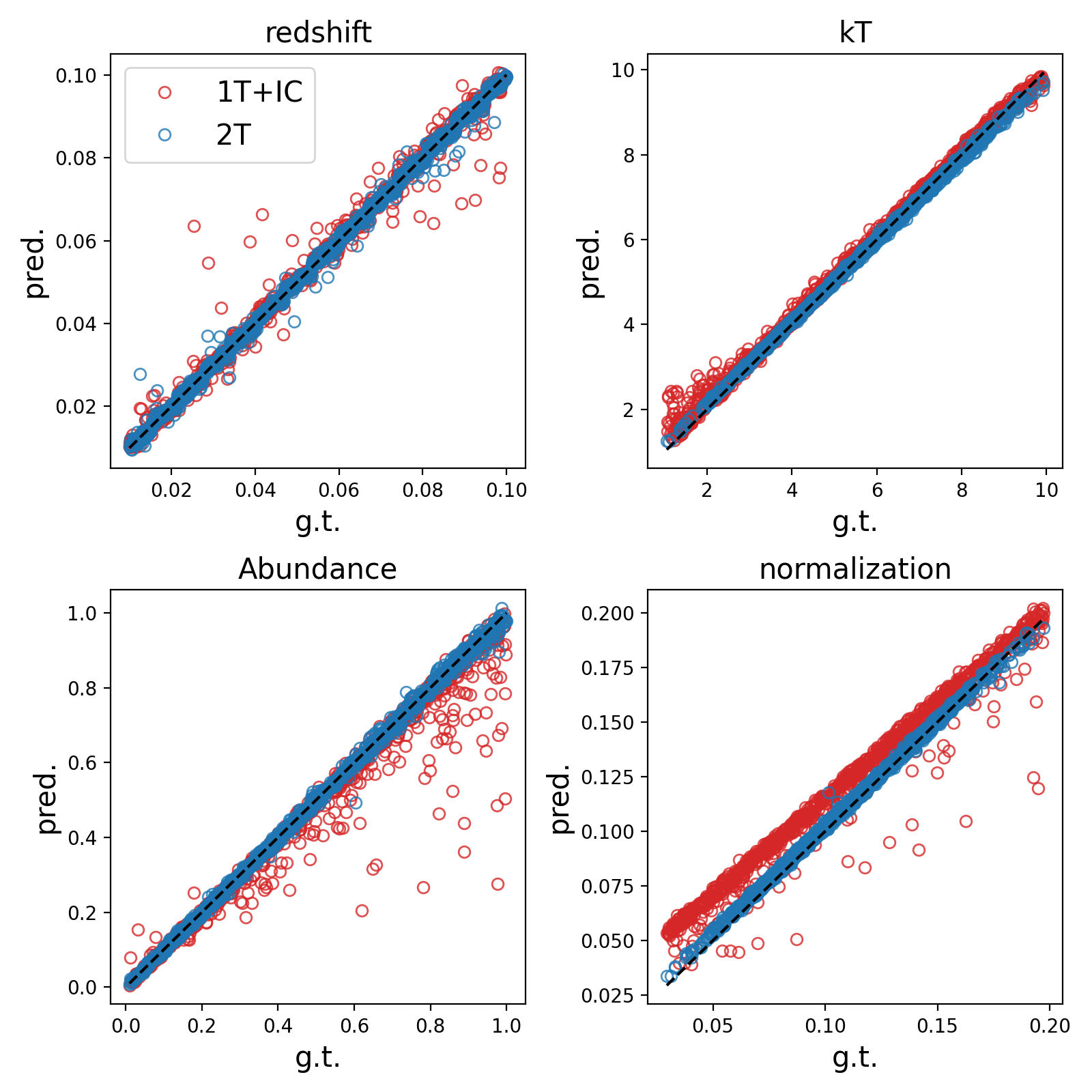}
    \caption{
        Latent representations as functions of thermal parameters $\theta = \{kT, Z, z, n\}$ of \texttt{apec}.
        The x-axis is the ground truth $\theta$; the y-axis is the constrained latent representation.
        True 1T+IC data are marked as red open circles, and true 2T data are marked as blue circles. The ground truth temperature and normalization of the 2T spectra is obtained by fitting the data with a 1T \texttt{apec} model. The black dashed line marks the 1:1 relation. 
    }
    \label{fig:code}
    \end{figure}

In addition to reconstructing the input spectra, CAE is trained to constrain the thermal parameters of the normal spectra using the latent space.
Although we never used the labels of the thermal parameters for testing or for the purpose of identifying abnormal spectra, it would be insightful to inspect how well CAE could extract the thermal parameters from the input spectra.
The resulting latent representations and the corresponding labels for two test sets, normal (2T) and abnormal (1T+IC), are shown in Fig.~\ref{fig:code}.
As expected, the latent variables of 2T are better constrained since it is the only data type seen by CAE.
Its ability to identify abnormal spectra during testing may be partly due to its inability to accurately predict the thermal parameters, preventing it from correctly reconstructing the input spectra. 
    \begin{figure*}
    \centering
    \includegraphics[width=15cm]{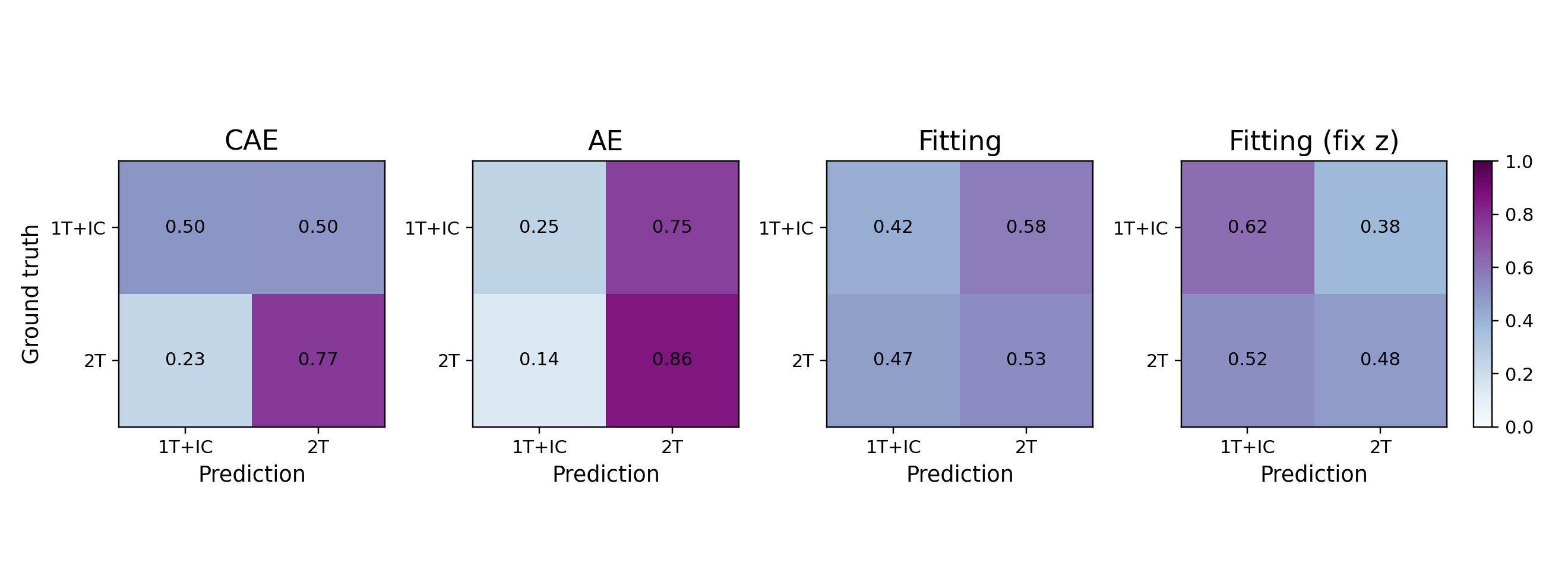}
    \caption{
        Confusion matrices of CAE, AE, and spectral-fitting methods.
        The y-axis labels the true spectra type, and the x-axis marks the predictions using these approaches.
    }
    \label{fig:bic_tradi}
    \end{figure*}

\subsection{Performance of traditional spectral fitting}

The similarity between 2T and 1T+IC spectra presents the biggest challenge for detecting IC signal in the spectra fitting.
Here we quantify the performance of traditional spectra fitting using a test dataset the same as that used in the performance evaluation for CAE and AE, including 1000 1T+IC and 1000 2T spectra.
We fit the spectra with both 1T+IC (\texttt{tbabs*(apec+po)}) and 2T (\texttt{tbabs*(apec+apec)}) models in XSPEC.
Each spectrum is binned with a minimum count of $100$ per energy bin.
The fitting range is set within $3$-$80$ keV, the same range used in the calculation of anomaly score.
The parameters of the background components except normalization are fixed at the values used in generating synthetic spectra.
To account for the statistical fluctuation, we let the normalizations of background components to vary within a $10\%$ range on the centroid of the input values.
The hydrogen column density $N_H$ is fixed at the input value of $8.58\times 10^{19}$ cm$^{-2}$ for both models and the power-law model has a photon index fixed at $2$.
The metallicity of two thermal components in 2T model are linked as they were in the simulated spectra.
All other parameters are left to vary freely.
The C-statistic is adopted to optimize the fitting.

We choose the model that gives a reduced C-statistic ($C/\mu$ where $\mu$ is the degrees of freedom) closer to $1$ as the preferred model (2T versus 1T+IC).
The resulting precision and recall are $0.54$ and $0.62$, respectively.
The BAcc of the spectral fitting is $0.55$, which is lower than CAE ($\rm{BAcc}=0.64$).
In reality, the ICM redshift is often unknown.
The performance of traditional spectra fitting is worse when we let the ICM redshift freely vary.
The full comparisons of each metric are shown in Table.~\ref{tab:metrics}.

\startlongtable
\begin{deluxetable}{cccc}

\tablecaption{Metrics evaluated from different models including CAE, AE, spectral fitting with $z$ fixed at the input redshift value, and fitting with $z$ freely varying.}

\startdata
\tablehead{\colhead{Model} & \colhead{Precision}& \colhead{Recall} & \colhead{BAcc}}
CAE & 0.69 & 0.50 & \textbf{0.64} \\
AE & 0.64 & 0.25 & 0.55 \\
fitting ($z$ fixed) & 0.54 & 0.62 & 0.55 \\
fitting ($z$ freed) & 0.47 & 0.42 & 0.47 \\
\enddata
\label{tab:metrics}
\end{deluxetable}

\subsection{Parameter distribution of performance measures}
We compare the performance of spectral fittings (with redshift fixed and thawed), CAE, and ordinary AE through their confusion matrix as shown in Fig.~\ref{fig:bic_tradi}, and their precision, recall, and BAcc, as listed in Table~\ref{tab:metrics}. Overall, CAE, with an improved BAcc, outperforms both traditional spectral fitting and ordinary AE. However, with a higher recall, spectral fitting is more capable of identifying IC cases among genuine IC spectra, while IC cases identified by CAE are more likely to be genuine IC spectra, as indicated by the higher precision of CAE. In other words, CAE and traditional spectral fitting are mutually complementary in detecting the IC signal.

\begin{figure*}
        \centering
        \includegraphics[width=15cm]{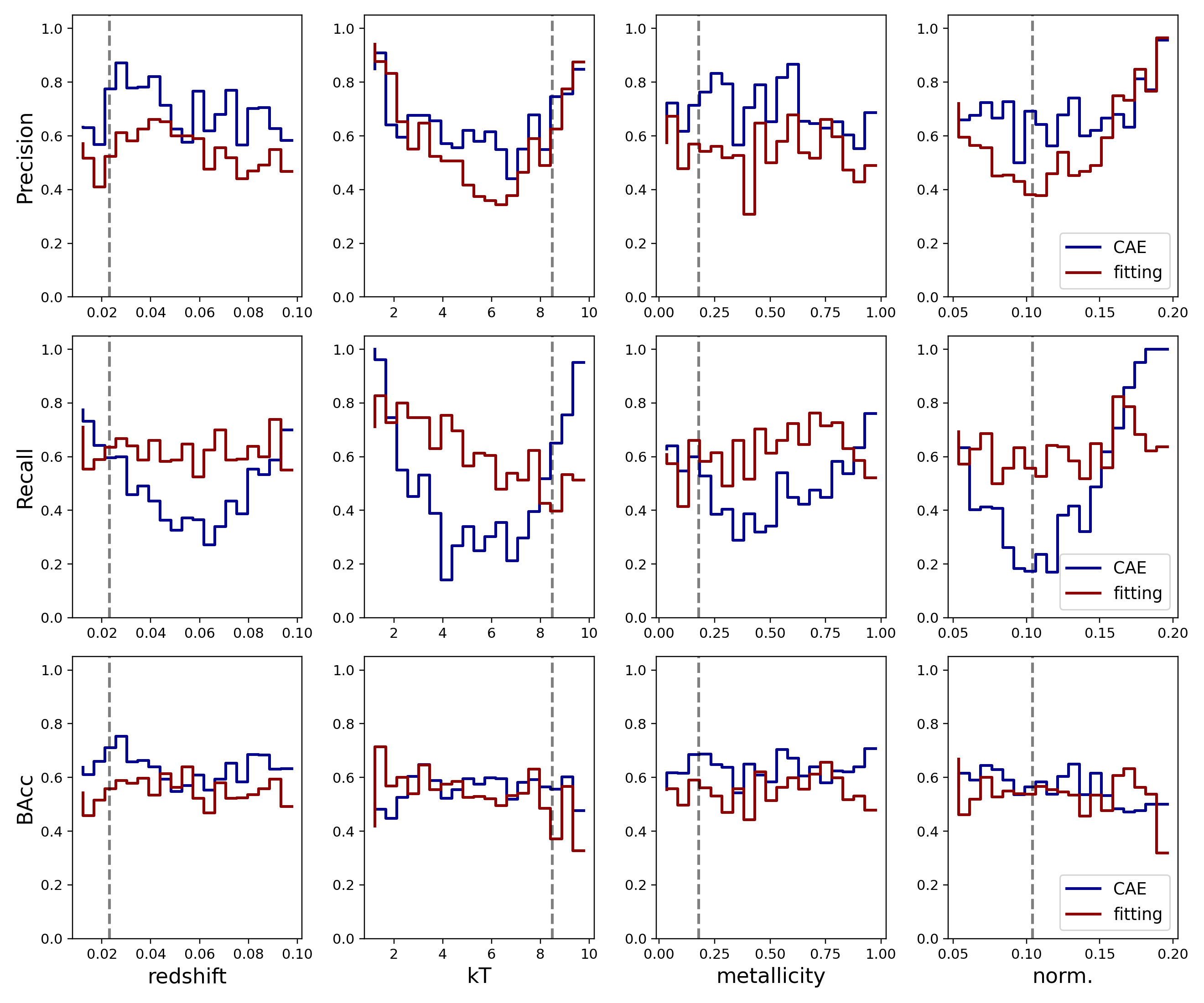}
        \caption{
            Performance measures (precision, recall, BAcc) as a function of thermal parameters for CAE (blue), and spectral-fitting results (red).
            The ground-truth labels left to right are redshift (z), temperature (kT), metallicity (Z), and normalization (norm.) of \texttt{apec}. 
            The thermal parameters of the Coma cluster, obtained from the spectral fitting, are indicated by the dashed lines, for which redshift is fixed at $0.023$ \citep{2015ApJ...800..139G}.
            Note that the redshift fixed in fitting is estimated from the velocity dispersion of member galaxies \citep{1999ApJS..125...35S}.
        }
        \label{fig:ae_cae}
    \end{figure*}

We inspect the distribution of thermal properties of the spectra in the test dataset as a function of recall, precision, and BAcc for CAE and spectral fitting (with $z$ fixed at the input redshift value), as shown in Fig.~\ref{fig:ae_cae}.
The overall performances, as indicated by BAcc, are insensitive to the thermal parameters for both CAE and fitting. Both approaches achieve higher precision for extreme (low and high) temperatures and high thermal normalization.
For lower-temperature systems, hard X-ray band would be less contaminated by the thermal emission. For the same metallicity/redshift/normalization, higher-temperature systems would have a higher flux in the energy band 3-80\,keV, allowing its thermal properties to be better determined.
Similarly, the high thermal normalization may allow their thermal parameters to be more easily constrained to anchor the thermal component for detecting the IC emission.
CAE achieves better recall for more extreme values of redshift, temperature, metallicity, and thermal normalization, whereas the recall of spectral fitting is not so sensitive to thermal properties. 

\section{Summary}
\label{sec:summary}

 To our knowledge, NuSTAR is, and will be in the near future, the primary observatory to focus hard X-ray. It is indispensable to maximize its science return using modern data analysis tools.
 We have presented an anomaly detection algorithm designed for searching inverse Compton (IC) emission using hard X-ray spectra of NuSTAR. 
 The algorithm is based on a conditional autoencoder (CAE) with latent representations trained to constrain the thermal parameters of the spectra. 
Synthetic NuSTAR spectra are used for training, validation, and testing, which are generated from two different models: two-temperature plasma model (2T) and a one-temperature model with an IC component (1T+IC). 
The training is based solely on the 2T dataset.
Meanwhile, the latent representations are constrained by the global thermal parameters of 2T.
The training converges within $300$ epochs, taking approximately an hour of GPU time.
The latent space shows that the thermal parameters of the spectra are successfully recovered for normal data sets.
This implies that as a byproduct of this algorithm, it can be used to measure thermal parameters for normal spectra.

We use the mean squared error to quantify the reconstruction accuracy.
As shown in Sec.~\ref{sec:results}, the reconstruction loss (MSE) of 2T test data set has a relatively narrow distribution, while that of 1T+IC displays a higher loss tail.
The distinguishable distributions of 1T+IC and 2T allow us to obtain a criterion to detect anomalous spectra (1T+IC).
Note that we adopted the same particular realization of the background model, based on the background of the Coma observation, for generating all the synthetic spectra.
Our results do not take into account systematic uncertainties in the background modelization of the observation for this initial work and it will be taken into account in future work.
When tested on the same data, the conditional autoencoder achieves an improvement in balanced accuracy compared to the traditional spectral fitting and ordinary autoencoder. 
Although CAE has higher precision, traditional spectral fitting has higher recall, that is, IC cases classified by CAE are more likely to be bonafide IC spectra, while bonafide IC spectra are less likely to be missed by spectral fitting, demonstrating their mutual complement in detecting IC signals.

\section*{Acknowledgements}

We would thank the University of Kentucky Center for Computational Sciences and Information Technology Services Research Computing for their support and use of the Lipscomb Compute Cluster and associated research computing resources. 
S.-C.\ L. and Y.\ S. were supported by Chandra X-ray Observatory grant GO1-22126X, NASA grant 80NSSC21K0714, and NSF grant 2107711.

This research has made use of data from the NuSTAR mission, a project led by the California Institute of Technology, managed by the Jet Propulsion Laboratory, and funded by the National Aeronautics and Space Administration. Data analysis was performed using the NuSTAR Data Analysis Software (NuSTARDAS), jointly developed by the ASI Science Data Center (SSDC, Italy) and the California Institute of Technology (USA).

\section*{Data Availability}

The data underlying this article will be shared on reasonable request to the corresponding author.

\bibliography{references}{}
\bibliographystyle{aasjournal}

\end{document}